# A 100 ps TOF Detection System for On-Line Range-Monitoring in Hadrontherapy

Sara Marcatili[1], Sébastien Curtoni[1], Denis Dauvergne[1], Ferid Haddad[2], Maxime Jacquet[1], Charbel Koumeir[2], Jean Michel Létang[3], Jayde Livingstone[1], Vincent Métivier[2], Laurent Gallin-Martel[1], Marie-Laure Gallin-Martel[1], Jean-François Muraz[1], Noel Servagent[2] and Étienne Testa[4]

*[1]Laboratoire de Physique Subatomique et Cosmologie (LPSC) and Grenoble-Alpes University, Grenoble, France. (Corresponding author e-mail: sara.marcatili@ lpsc.in2p3.fr).*
*[2]SUBATECH laboratory, IMT Atlantique and Nantes University, Nantes, France.*
*[3]Lyon University, CREATIS, CNRS UMR5220, Inserm U1044, INSA-Lyon, Univ. Lyon 1, France.*
*[4]Lyon University, Claude Bernard Lyon1University, CNRS/IN2P3, IP2I Lyon, F-69622, Villeurbanne, France.*

***Abstract*—The accuracy of hadrontherapy treatment is currently limited by ion-range uncertainties. In order to fully exploit the potential of this technique, we propose the development of a novel system for online control of particle therapy, based on TOF-resolved (time-of-flight) Prompt Gamma (PG) imaging with 100 ps time resolution. Our aim is to detect a possible deviation of the proton range with respect to treatment planning within the first few irradiation spots at the beginning of the session.**

**The system consists of a diamond-based beam hodoscope for single proton tagging, operated in time coincidence with one or more gamma detectors placed downstream of the patient. The TOF between the proton time of arrival in the hodoscope and the PG detection time provides an indirect measurement of the proton range in the patient with a precision strictly related to the system time resolution. With a single ~38 cm$^3$ BaF$_2$ detector placed at 15 cm from a heterogeneous PMMA target, we obtained a coincidence time resolution of 101 ps (rms). This system allowed us to measure the thickness and position of an air cavity within a PMMA target, and the associated proton range shift: a 3 mm shift can be detected at 2σ confidence level within a single large irradiation spot (~10$^8$ protons).**

**We are currently conceiving a multi-channel PG timing [1,2] detector with 3D target coverage. Each pixel of about 5×5 mm$^2$ detection surface will provide the PG detection time and its hit position, that can be used to reconstruct the longitudinal distribution of PG vertices in the patient. The number of PG detected in each channel is used to reconstruct the vertex in the transverse plane. Our approach does not require collimation and allows to dramatically increase the detection efficiency. Since both signal detection and background rejection are based on TOF, the constraints on energy resolution can be relaxed to further improve time resolution. The pixel detector technology is currently under test and will be based on Cherenkov radiators coupled to Silicon Photomultipliers (SiPM).**

## I. Introduction

In hadrontherapy, tumors are treated by means of light ion beams (mainly of protons and $^{12}$C). In principle, hadrontherapy treatments offer a highly conformal tumor coverage thanks to the typical dose depth profiles of ions presenting a maximum at the end of the ion range. However, uncertainties in tissues composition, patient positioning or physiological modifications of patient anatomy may result in a significant deviation of the ion range with respect to treatment planning, thus limiting *de facto* the precision achievable with this technique. In this context, the use of an online range monitoring system is crucial to promptly detect any discrepancy with the planned treatment, and possibly interrupt the session at its very beginning.

Several methods have already been developed to measure the ion range online or between treatment sessions [3,4,5]. We propose a novel system for online control of particle therapy, based on TOF-resolved PG imaging with 100 ps time resolution. The system is composed by one or more small size (<1 cm$^3$) gamma detectors closely surrounding the target region and read in time coincidence with a diamond-based beam hodoscope developed at LPSC [6]. The time occurred between a proton time of arrival in the hodoscope and a PG detection time corresponds to the travel time of the ion inside the patient plus the PG TOF. With the spatial position of gamma detectors known *a priori*, this time can be used to analytically determine the ion range with a precision depending on the system time resolution and on the spatial extent of the gamma detectors.

Our system presents several elements of novelties. The absence of a collimation system together with a close detection distance, result in a high total detection efficiency compensating the relatively low detection efficiency of each pixel. In addition, since both signal detection and background rejection are based on TOF, the constraints on energy resolution can be relaxed to further improve time resolution.

Our objective is to operate this system in a single proton counting regime during the first few irradiation spots probing the whole treatment: this approach would allow performing an analytical proton range reconstruction thus enabling clinicians to react in real time in case of discrepancy with respect to treatment planning.

## II. TOF BASED RANGE MONITORING WITH A SINGLE DETECTOR

We already developed a single channel version of the proposed detector to demonstrate the feasibility of our approach. Two PMMA targets of 0.5 (T1) and 10 cm (T2) thickness were irradiated with 68 MeV protons at the ARRONAX facility in Nantes, France [7]. T1 was fixed at 18.5 cm from a 5×5×0.5 $mm^3$ single-crystal diamond detector placed in the beam axis. T2 was progressively moved away from T1 (at 25, 35, 50, 70, 100, 150 mm) in order to simulate the presence of a variable thickness air cavity. Three large size (~38 $cm^3$) gamma detectors were arranged at 15 cm distance from the center of the T1, and read independently with fast photomultipliers: two $LaBr_3$ crystals placed at 90° and 120° with respect to the beam axis, and a $BaF_2$ scintillator at 120°. All signals were acquired with a 3.2 GHz multi-channel digital sampler, namely the Wavecatcher [8].

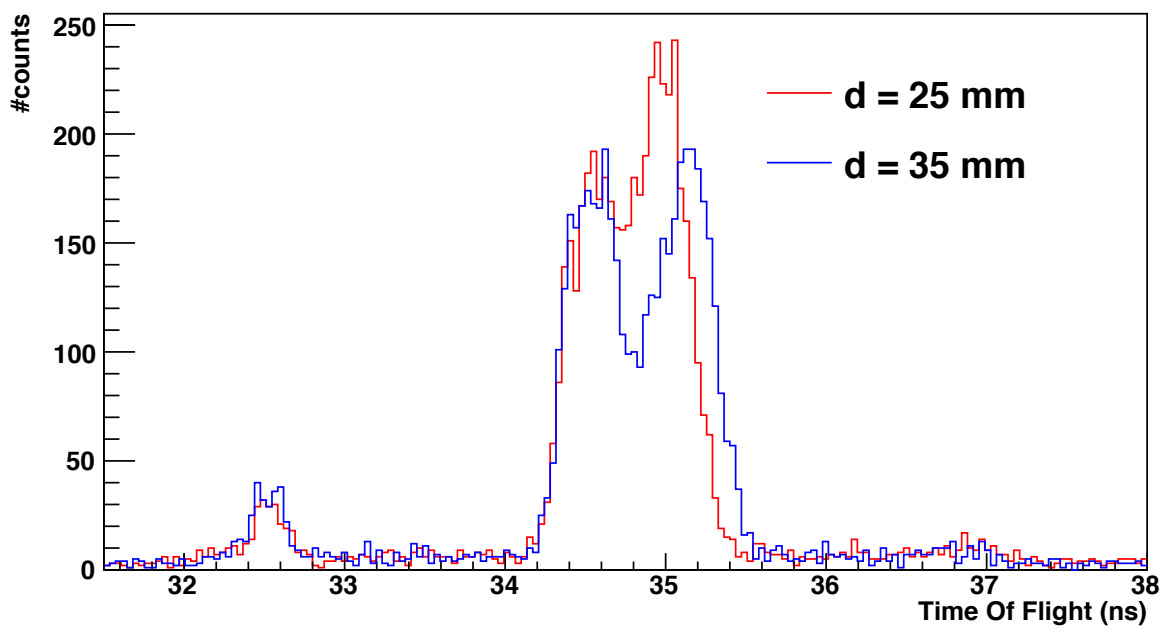

Fig. 1. PG TOF spectra for two different positions of T2 (d=25 mm and 35 mm) obtained with a $BaF_2$ detector placed at 120° with respect to the beam axis.

Figure 1 shows the TOF spectra obtained with the $BaF_2$ detector at 120° in the case of a target-to-target distance of 25 and 35 mm. In the spectra, the first peak corresponds to PGs generated in the hodoscope and provides a spatial reference for patient positioning along the beam axis; its dispersion is also a measurement of the system Coincidence Time Resolution (CTR). A CTR of 101 ps (rms) was obtained for the $BaF_2$. The second peak corresponds to PGs generated in T1, while the third shifts accordingly to T2 position. We measured the deviation in proton range for different positions of T2 with a notion of T1-to-T2 time distance defined as follows. The PG TOF spectra are used to build their normalized integral function: each peak in the original spectra corresponds to an inflection point in the integral function. The T1-to-T2 distance is then calculated applying two fixed thresholds at the second and third inflection points in the integral function and measuring the relative time delay. This approach allows minimizing the impact of statistical fluctuations in the PG timing spectra, taking advantage of the fast convergence of the integral function. Fig. 2 shows the linear relation found between the measured and the actual T1-to-T2 distances. Experimental errors are within the point size.

### *A. Sensitivity study*

The limited extent of the experimental errors in Fig. 2 suggests that the sensitivity of our approach is much better than that explored experimentally, possibly allowing to detect range-shifts below 25 mm. We therefore investigated the technique sensitivity through Monte-Carlo (MC) simulations. Using Geant4.10.5 we reproduced the experimental set-up assuming T2 shifts from 25 to 30 mm in 1 mm steps, and progressively reducing the simulation statistics from $2.6\times10^9$ to $10^8$ primary protons. For example, using a single 38 $cm^3$ $LaBr_3$ detector, a range shift of 1-2 mm can be detected at 1σ confidence level within a single irradiation spot of $10^8$ primary protons. If a significance level of 2σ is targeted, a 3-4 mm shift would be detectable under the same conditions.

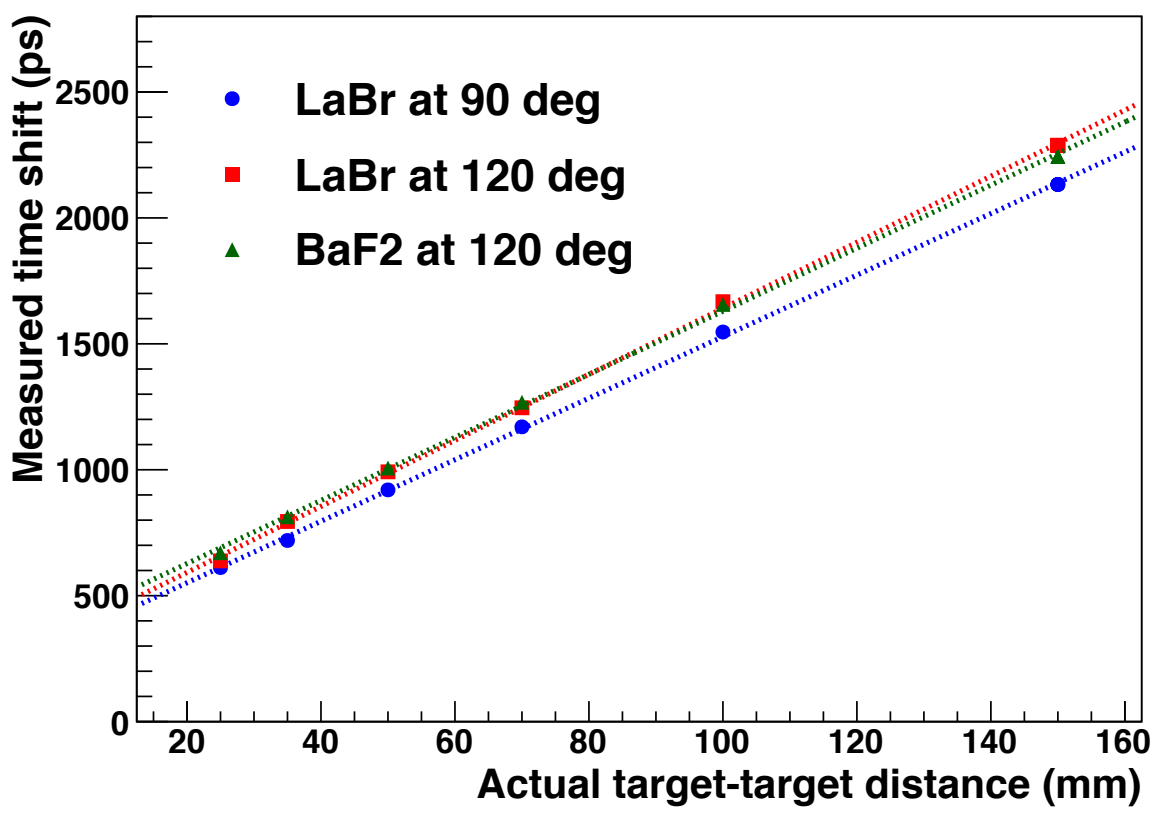

Fig. 2. Measured time shifts as a function of T1-to-T2 distance obtained with three different gamma detectors. Error bars are within the symbols.

## III. CONCEPTION OF A TOF BASED 3D IMAGING ARRAY

We are currently conceiving a PG-timing multi-channel system to achieve 3D target coverage, namely the TIARA (Time-of-flight Imaging ARrAy) detector. A similar detection volume as the detector already tested will be split in ~30 pixels surrounding the target. The block detector technology is under investigation: we are currently testing 5×5×10-20 $mm^3$ $PbF_2$ Cherenkov radiators coupled to Silicon Photomultipliers. The use of Cherenkov radiators is advantageous for two main reasons: first, the Cherenkov process is inherently faster than the scintillation one, allowing in principle to achieve better time resolutions; secondly, Cherenkov radiators are typically high density materials, making it possible to detect high energy gamma rays with relatively small volumes keeping the detector compact.

Our aim is to achieve a maximum CTR of 100 ps (rms) for each pixel separately. A very limited energy resolution is needed to increase the efficacy of background rejection (we apply a 3-8 MeV selection) also obtained from TOF. Each gamma detector will provide the PG hit position and its TOF with respect to the hodoscope. The PG hit coordinates and its TOF geometrically constrain the PG vertex coordinates, allowing their analytical reconstruction on an event-by-event basis.

### *A. Monte-Carlo simulation*

In order to assess the capability of this system to detect proton range shifts of few millimeters we have simulated with Geant4 its response in the case of a spherical arrangement around a head phantom (10 cm overall radius, 7 mm skull) irradiated with a 100 MeV proton pencil beam. The detector array was composed of 30 pixels placed on the head surface and presenting a detection efficiency of 20% and a CTR of 100 ps (rms). The beam energy was varied between 100 and 105 MeV in steps of 1 MeV in order to generate a shift in the proton range. The PG vertex profile along the beam axis has been reconstructed for each simulated energy.

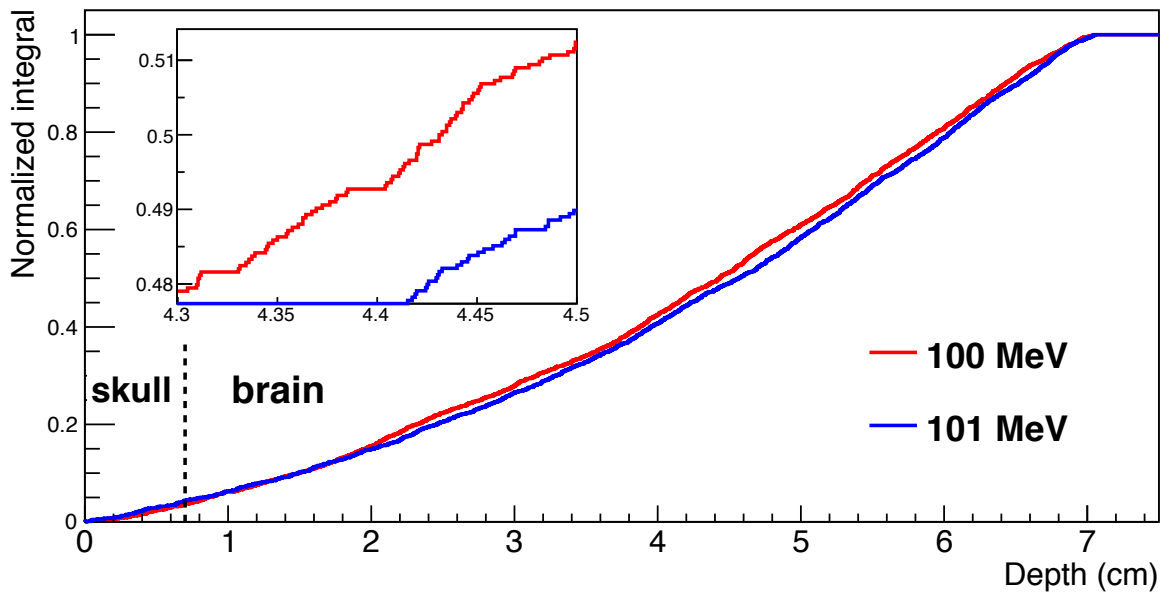

Fig. 3. Integral curve of reconstructed PG longitudinal vertex distributions obtained with the head phantom and $10^8$ incident protons. The inset shows the measurable separation between the curves.

Fig. 3 shows the integral function of the PG reconstructed profiles in the case of 100 and 101 MeV proton energy; the curves correspond to $10^8$ incident protons impinging on the head at Depth=0. Bragg peaks fall at 6.86 and 7.01 cm for 100 and 101 MeV proton beams respectively. The curves are well separated all along the proton range within the phantom, as highlighted in the inset, and their distance is measurable.

Integrating the PG vertex profiles allows reducing the statistical fluctuations of the distributions so as to highlight the shift between the curves.

### B. *Experimental feasibility study*

A proof-of-principle experiment was performed at ARRONAX cyclotron to show the feasibility of Cherenkov-based PG detection. A PMMA target of 5 cm depth was irradiated with 68 MeV protons. The beam intensity was reduced down to 1.5 protons per bunch at the hodoscope level.

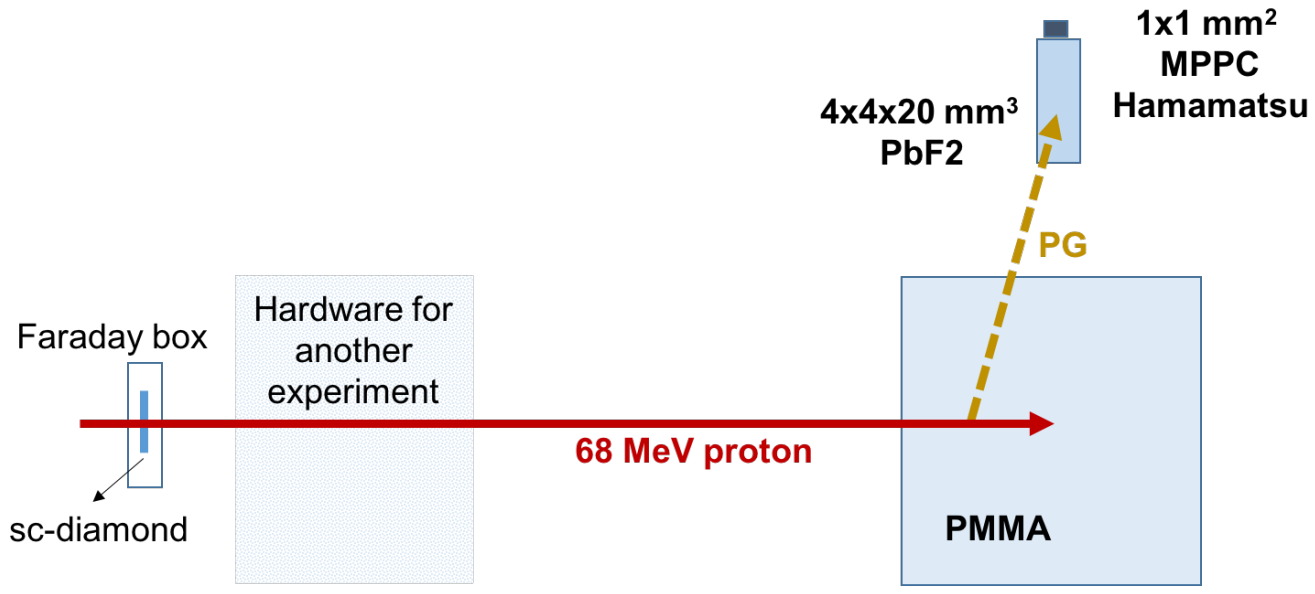

Fig. 4. Experimental set-up for PG TOF measurement with a Cherenkov radiator.

Upstream a single-crystal diamond detector was used as beam hodoscope to provide the proton time stamp for the PG TOF measurement. Behind the diamond detector, hardware of various thickness for another experiment was arranged at ~20 cm from the target; the overall attenuation results in a proton absorption within approximately 2.9 cm in the PMMA target. Downstream, a 4×4×20 mm$^3$ $PbF_2$ Cherenkov radiator wrapped in white Teflon, is read by a 1×1 mm$^2$ SiPM detector from Hamamatsu (S12260 series) in a ceramic package. The crystal and the SiPM are coupled using optical grease (BC630 from Saint-Gobain). This module provides the PG detection time for the PG TOF measurement: SiPM signals were amplified using the C12332-01 evaluation circuit from Hamamatsu and acquired with a 4 GS/s oscilloscope for an offline analysis.

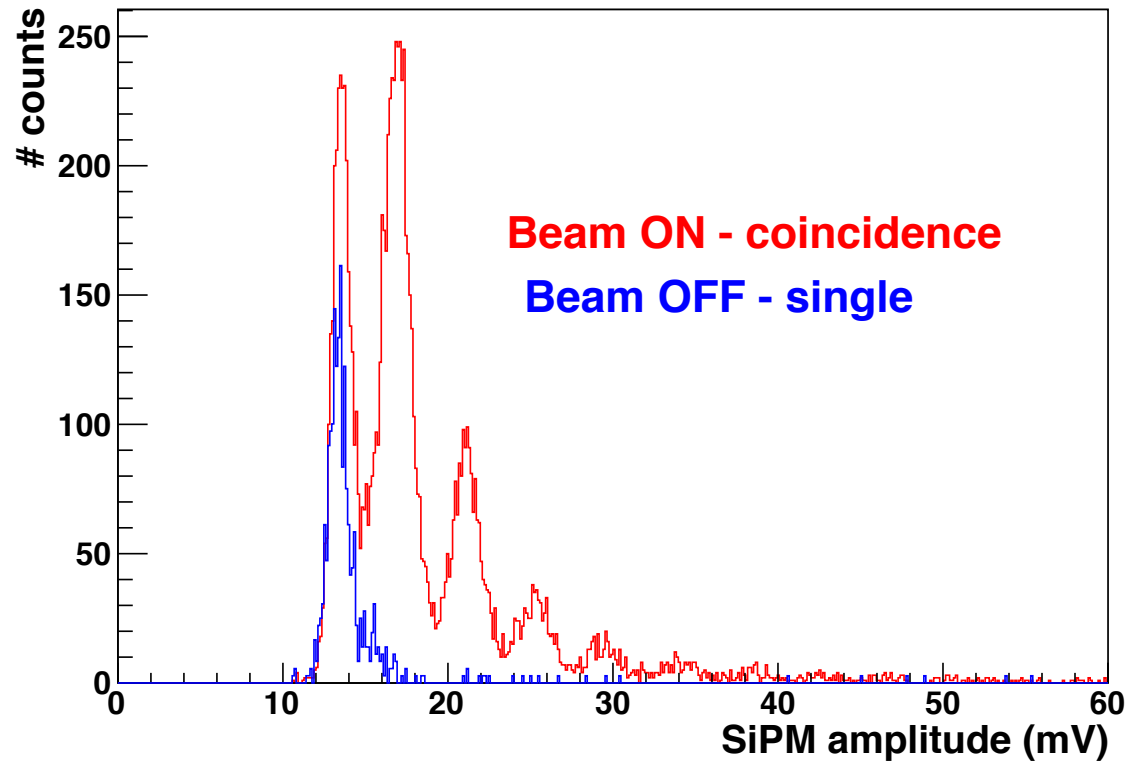

Fig. 5. Amplitude spectra obtained with the $PbF_2$ crystal read out by a SiPM, under beam-on (red) and beam-off conditions. The beam-off acquisition has been performed just after the beam-on one.

Fig. 5 shows the PG energy spectra obtained with the $PbF_2$ crystal; a threshold between 3 and 4 photoelectrons was applied to the SiPM. Data in red represent the spectrum acquired in time coincidence with the diamond detector under beam-on conditions: up to 8-9 Cherenkov photons could be detected with this non-optimized set-up (in terms of SiPM wavelength sensibility, detection surface and optical coupling). Data in blue corresponds to the amplitude spectrum at beam off (right after irradiation), with the background mainly consisting of gamma rays from the de-excitations of activated materials in the beam line and vault. The two spectra are normalized in terms of equal acquisition times.

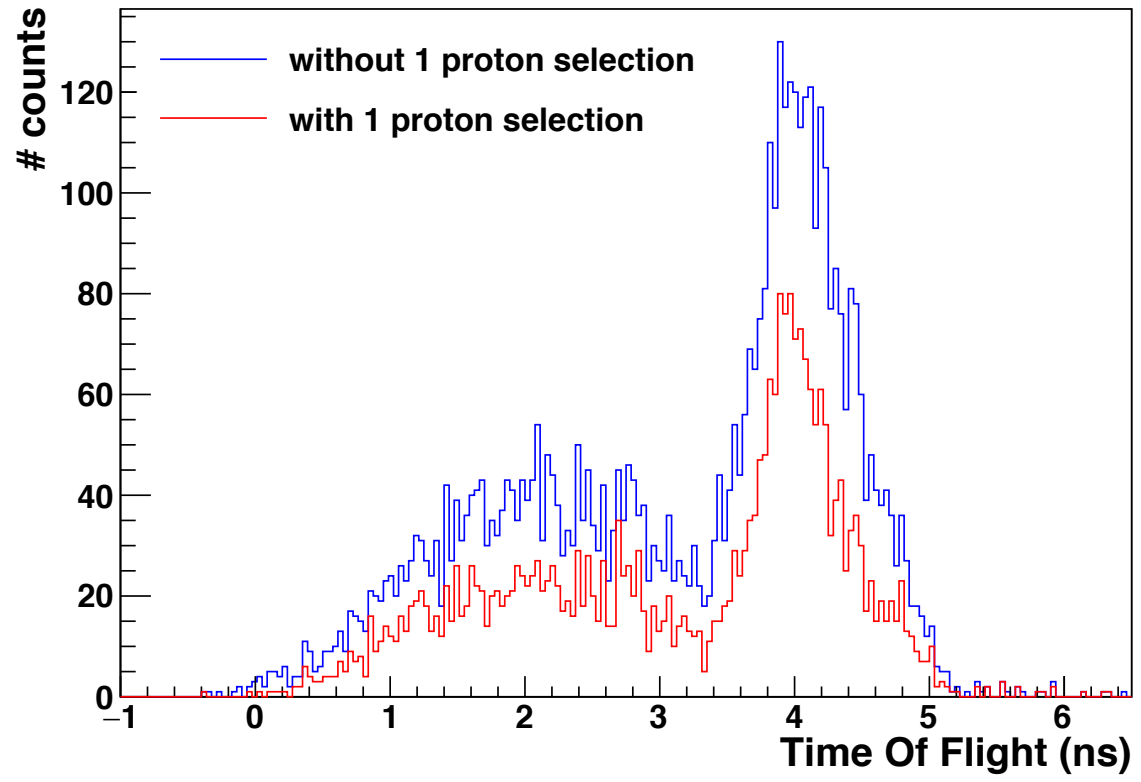

Fig. 6. PG timing spectra obtained from all detected events (blue) and from single-proton events (red).

All events acquired with the threshold at 3-4 photoelectrons were used to build the TOF distributions in Fig. 6. Despite using a threshold lower than the level of background in the vault we were able to achieve a good signal to noise ratio thanks to the narrow coincidence window achievable with our set-up.

Two main regions are recognizable in the PG timing spectra of Fig. 6: the first part of the distribution corresponds to PG generated in the hardware from the experiment placed upstream, while the sharper peak corresponds to PG generated in the PMMA target. The time distance of ~2 ns between these two regions is in agreement with the proton TOF between the two experimental set-up (placed at ~20 cm distance).

Even if a direct measurement of the system CTR was not possible for this experiment, it is possible to safely assume, from comparison with previous Monte-Carlo simulations performed under similar conditions, that CTR was of the order of few hundreds of ps. From the comparison of the two spectra in Fig. 6, it is also possible to observe that, when only events for which a single hit has occurred in the hodoscope are selected (red curve), the width of the PG distribution narrows by ~15% as a consequence of the achievable improved time resolution.

## IV. Conclusions

We are developing an online monitoring system for hadrontherapy exclusively based on TOF detection. This method may provide real-time information thanks to an analytical PG vertex reconstruction. In addition, this approach makes it possible an extreme simplification of the detector hardware and possibly a high detection efficiency thanks to the absence of any collimation system and to the close detection distance. We have proven the feasibility of our approach with a single channel, large volume $BaF_2$ detector obtaining a 101 ps (rms) coincidence time resolution, and with a small size $PbF_2$ Cherenkov radiator.

We are currently conceiving a multi-channel, Cherenkov radiator-based detector in order to achieve 3D target coverage.

## V. Conclusions

The authors would like to thank ITMO-Cancer (CLaRyS-UFT project). This work is carried out in the frame of Labex PRIMES (ANR-11-LABX-0063). Part of this work was performed within the framework of the EU Horizon 2020 project RIA-ENSAR2 (654 002) and is partly supported by the French National Agency for Research called "Investissements d'Avenir", Equipex Arronax-Plus ANR-11-EQPX-0004, Labex IRON ANR-11-LABX-18-01 and ANR-16-IDEX-0007.